\documentclass[
  reprint,                
  amsmath,amssymb,
  aps,                    
  pra                     
]{revtex4-2}

\usepackage{graphicx}     
\usepackage{dcolumn}      
\usepackage{bm}           
\usepackage{hyperref}     
\usepackage{xcolor}       
\usepackage{physics}      
\usepackage{soul}
\setstcolor{blue}
\usepackage{subcaption}
\usepackage{float}
\usepackage{placeins}
\newcommand{\transp}{\mathsf T}

\begin{document}

\preprint{APS/123-QED}

\title{Complex frequency-dependent quadrature squeezing in semiconductor lasers}

\author{Daniele Nello}
\affiliation{Laboratoire PhLAM (UMR 8523), Universit\'e de Lille, F-59000 Lille, France}
\affiliation{Dipartimento di Elettronica e Telecomunicazioni, Politecnico di Torino, Torino 10129, Italy}
\email{daniele.nello@univ-lille.fr}
\email{daniele.nello@polito.it}

\author{Giuseppe Patera}
\affiliation{Laboratoire PhLAM (UMR 8523), Universit\'e de Lille, F-59000 Lille, France}

\author{Lorenzo Columbo}
\affiliation{Dipartimento di Elettronica e Telecomunicazioni, Politecnico di Torino, Torino 10129, Italy}

\date{\today}

\begin{abstract}
We present a comprehensive study of quadrature squeezing in a quantum well laser based on a fully quantum Langevin approach. We compute the frequency-resolved squeezing map of the laser field and identify optimal squeezing curves, revealing, for the first time to our knowledge, both frequency-dependent squeezing and complex or hidden squeezing in a semiconductor laser. We further analyse the role of the linewidth enhancement factor ($\alpha$-factor) in the emergence of these features. Our results establish semiconductor lasers as a platform for the generation of non-classical light and open new perspectives for their application in quantum communication and sensing.
\end{abstract}

\maketitle



\section{Introduction}



Lasers are one of the most important inventions of the XX century, which, although being a mature commercial technology, witness a continuous stream of innovations and technological progress. Driven by the need for compact, low-power, and chip-compatible quantum light sources, such active photonic devices are emerging as alternative platforms for the generation of squeezed light, with the potential to produce bright squeezed states in integrated architectures.

Squeezed states have promising applications in various experimental and technological platforms. In particular, their below-shot-noise characteristics can be leveraged in gravitational interferometry~\cite{grav_interferometry,ligo_collaboration,freq-dep}, and biological sensing~\cite{biolog1,biolog2}. Moreover, squeezed states have been shown to be able to overcome the performance of coherent states in secure quantum communications even with a few dB of squeezing, with special reference to continuous-variable quantum key distribution protocols~\cite{QKD1,QKD2,QKD3}.

Early experimental work showed that injecting a sub-Poissonian electron stream, achievable via high-impedance current suppression in a regime dubbed \emph{quiet pumping}, produces a relative intensity noise spectrum below the shot-noise level at low frequencies~\cite{yamamoto1,yamamoto2}. Since the phase of a free-running laser is not fixed but undergoes diffusive motion, any noise reduction is naturally expected to emerge in the amplitude quadrature, which is directly related to photon-number fluctuations~\cite{davidovich}. Accordingly, previous studies of squeezing in lasers have mainly focused on amplitude spectrum.

Yet this picture captures only part of the quantum-noise structure of semiconductor lasers. In this work, we develop a general framework that goes beyond the amplitude quadrature and gives access to the full quadrature-noise structure of the emitted field. To this aim, we develop a quantum Langevin input-output treatment~\cite{quantumLangevin} of the linearized fluctuations of a semiconductor laser. Our analysis is based on a microscopic model of a quantum-well active material~\cite{ChowKochSargent1994} and retains the full dynamics of the active medium, avoiding adiabatic elimination. This differs from previous approaches, which have mostly relied on semi-classical methods where radiation was described in terms of classical observables obeying Maxwell equations~\cite{nillson,coldren,elsasser,VCSEL1}, or on analyses restricted to the amplitude-noise spectrum~\cite{travagnin}.

We consider the full set of coupled quantum fluctuations of the optical field and carrier population, allowing us to construct the spectral covariance matrix of the emitted light. This reveals quantum-noise properties that cannot be inferred from intensity-noise measurements alone. In particular, the quadrature exhibiting maximum squeezing generally depends on frequency. While the amplitude quadrature remains optimal at low frequencies, the optimal squeezing quadrature rotates at higher frequencies, extending the sub-shot-noise bandwidth beyond that predicted by amplitude squeezing alone. This effect is largely driven by the Henry linewidth-enhancement factor~\cite{alpha_henry}, which couples amplitude and phase fluctuations in semiconductor lasers. The latter is largely irrelevant to RIN measurements but is essential for the emergence of frequency-dependent squeezing and, as we show below, hidden squeezing.

The spectral covariance matrix also contains information inaccessible to standard homodyne detection. In particular, its imaginary part encodes correlations between optical sidebands, giving rise to the so-called hidden~\cite{barbosa,barbosa2,Gouzien2023,Dioum2025HiddenCorrelations} or complex~\cite{buchmann} squeezing. Such correlations can be accessed through generalized detection schemes, including synodyne detection~\cite{buchmann}, resonator detection~\cite{barbosa,barbosa2}, and interferometers with memory effect~\cite{Dioum2024}. To the best of our knowledge, this constitutes the first prediction of hidden squeezing in a semiconductor laser. More broadly, these results suggest new opportunities for quantum sensing and metrology, similarly to what has been proposed in optomechanical force-sensing protocols~\cite{buchmann}.

\section{The linearized quantum Langevin equations and the covariance matrix}
We start our analysis with the quantum-mechanical model for the quantum-well laser, introduced in \cite{ChowKochSargent1994} to justify the stochastic rate equations for the laser from its underlying microscopic dynamics. We note, however, that the methodologies presented here are completely general and can be applied also to other typologies of semiconductor lasers, such as quantum dot or QCLs, of much interest in the community.

The system of macroscopic quantum Langevin equations for the creation/annihilation operator of the photon field inside the laser cavity $A$ and the number of electron-hole pairs in the quantum well $N$ reads
\begin{equation}
    \begin{cases}        
    \dot{A}=-\frac{1}{2\tau_p}A+(1-i\alpha_H)\frac{g(N)}{2} A+F_{A} \\
    \dot{N}=\Lambda-R_{\mathrm{sp}}-\gamma_{\mathrm{nr}}N-g(N)A^\dag A+F_N.
    \end{cases}
\end{equation}
Here we indicated $\tau_p$ as the photon lifetime, $\Lambda$ as the electronic pumping rate, parametrized as $\Lambda=(1+a)\Lambda_{\mathrm{th}}$, where $\Lambda_{\mathrm{th}}$ is the threshold pumping current, $R_{\mathrm{sp}}$ is the spontaneous emission rate, $\gamma_{\mathrm{nr}}$ is the non-radiative decay rate per unity of carriers and finally $g(N)$ represents the gain, with $\alpha_H$ the Henry alpha factor. The stochastic part features a couple of correlated noise terms $F_A$ and $F_N$, whose correlations are reported in the Supplementary Material. 

Rather than taking the semiclassical limit by identifying the operators with commuting classical variables and arriving at the widely used stochastic rate equations for the laser dynamics \cite {coldren}, we keep our system of Quantum Langevin equations, and we linearise this system of equations around the semiclassical CW stationary solution \cite{gouzien}, approximating the field annihilation operator and the number of carriers as 
\begin{eqnarray}
    A\simeq (A_0^s+\delta A)e^{i\nu_L t}\\
    N\simeq N_{st}+\delta N,
\end{eqnarray}
where $A_0^s=\sqrt{N_p^{st}}$ is the square root of the stationary number of photons in the cavity and $\nu_L$ is the laser frequency, with corrections due to the alpha factor.

This yields the following linear system of equations for the quadratures \footnote{Here we define the quadrature variations in terms of $\delta A$ and $\delta A^\dag$ with the convention $\delta X=\frac{\delta A+\delta A^\dag}{2}$ and $\delta Y=\frac{\delta A-\delta A^\dag}{2i}$} variation $\delta X, \delta Y, \delta N$ with respect to stationary values
\begin{equation}
\begin{pmatrix}
\delta \dot{X} \\
\delta \dot{Y} \\
\delta \dot{N}
\end{pmatrix}
=
\begin{pmatrix}
0 & 0 & \gamma_{XN} \\
0 &0& -\gamma_{YN} \\
-\gamma_{NX} & 0 & -\gamma_{NN}
\end{pmatrix}
\begin{pmatrix}
\delta X \\
\delta Y \\
\delta N
\end{pmatrix}
+
\begin{pmatrix}
F_X \\
F_Y \\
F_N
\end{pmatrix},
\end{equation}
\vspace{0.5 cm} 
where their associated noise terms are $F_X,F_Y,F_N$, 
$\gamma_{XN}=A_0^s/2 \frac{\partial g(N)}{\partial N}\Big|_{\mathrm{st}}$,
$\gamma_{YN}=A_0^s \alpha_H/2  \frac{\partial g(N)}{\partial N}\Big|_{\mathrm{st}}$,
$\gamma_{NX}=2 A_0^s g(N_{\mathrm{st}})$,$\gamma_{NN}=\frac{\partial \Gamma(N)}{\partial N}\Big|_{\mathrm{st}}+(A_O^s)^2 \frac{\partial g(N)}{\partial N}\Big|_{\mathrm{st}}$, and $\Gamma(N)=\gamma_{\mathrm{nr}}N+R_{\mathrm{sp}}$. We represent this in the compact form
\begin{equation}\label{nine}
\frac{d}{dt}\begin{pmatrix}
R_{\mathrm{opt}} \\
\hline
R_{\mathrm{car}}
\end{pmatrix}=\begin{pmatrix}
        0_{2*2} & \vline B \\ \hline
        C & \vline D
\end{pmatrix}\begin{pmatrix}
R_{\mathrm{opt}} \\
\hline
R_{\mathrm{car}}
\end{pmatrix}+\begin{pmatrix}
\mathcal{N}_{\mathrm{opt}}+\mathcal{N}_{P} \\
\hline
\mathcal{N}_{\mathrm{car}}
\end{pmatrix}
\end{equation}
separating the dynamics of the degrees of freedom relative to the optical quadratures ($R_{\mathrm{opt}}=(\delta X,\delta Y)^\transp$) and the carrier variables, $R_{\mathrm{car}}=\delta N$. With this convention, $B=(\gamma_{XN},-\gamma_{YN})^\transp$, $C=(-\gamma_{NX}, 0)$ and $D=-\gamma_{NN}$. Note that here, apart from the input optical noise $\mathcal{N}_{\mathrm{opt}}=-\sqrt{\kappa}R_{\mathrm{opt},\mathrm{in}}$, $R_{\mathrm{opt},\mathrm{in}}=(X_{\mathrm{in}},Y_{\mathrm{in}})^\transp$, we have a further source of noise in the quadratures $\mathcal{N}_{P}=(F_{X_P},F_{Y_P})^\transp$ due to the polarisation, which we have eliminated adiabatically, and that is correlated with $\mathcal{N}_{\mathrm{car}}=F_N$ \cite{ChowKochSargent1994}, in such a way that $(F_X,F_Y)^\transp=\mathcal{N}_{\mathrm{opt}}+\mathcal{N}_{P}$.

This system of equations admits a direct solution in Fourier domain~ \footnote{The notation used in this work for the Fourier transform is \begin{equation*} R(\omega)=\int_{-\infty}^{+\infty}\frac{\mathrm{d}t}{\sqrt{2\pi}}\,\mathrm{e}^{i \omega t} R(t)\end{equation*} and \begin{equation*} R(t)=\int_{-\infty}^{+\infty}\frac{\mathrm{d}\omega}{\sqrt{2\pi}}\,\mathrm{e}^{-i \omega t} R(\omega)\end{equation*}}. Using
the input-output relations in the form \cite{gardinercollett},
$R_{\mathrm{opt},\mathrm{out}}=R_{\mathrm{opt},\mathrm{in}}+\sqrt{\kappa}R_{\mathrm{opt}}$, with $\kappa=1/\tau_p$, this leads to the spectral form of the output quadratures, which has contributions from the three different channels 
\begin{equation}    
R_{\mathrm{opt},\mathrm{out}}(\omega)
=
S_{\mathrm{opt}}(\omega) R_{\mathrm{opt},\mathrm{in}}(\omega)+
S_{\mathrm{car}}(\omega)\mathcal{N}_{\mathrm{car}}+S_{P}(\omega) \mathcal{N}_{P},
\end{equation}
with their spectral scattering forms
\begin{align}  
&S_{\mathrm{opt}}(\omega)=\mathbf{1}_2-\kappa[j\omega\mathbf{1}_2-B(j\omega\mathbf{1}_1-D)^{-1}C]^{-1}\\
&S_{\mathrm{car}}(\omega)=\sqrt{\kappa}[j\omega\mathbf{1}_2-B(j\omega\mathbf{1}_1-D)^{-1}C]^{-1}B(j\omega\mathbf{1}_1-D)^{-1}\\
&S_P(\omega)=\sqrt{\kappa}[j\omega\mathbf{1}_2-B(j\omega\mathbf{1}_1-D)^{-1}C]^{-1}
\end{align}
from the solution of the system of Equations \ref{nine}. We consider a vacuum input, or zero-temperature bath, such that $\sigma_{\mathrm{opt},\mathrm{in}}(\omega)=\frac{1}{4\sqrt{2\pi}}\mathbb{I}_2$, which is $\omega$-independent. The spectral covariance matrix is defined as $\Sigma_{\mathrm{opt},\mathrm{out}}=\frac{1}{2}\langle R_{\mathrm{opt},\mathrm{out}}(\omega)R_{\mathrm{opt},\mathrm{out}}^T(-\omega)+(R_{\mathrm{opt},\mathrm{out}}(-\omega)R^T_{\mathrm{opt},\mathrm{out}}(\omega))^T\rangle$, which in our single mode case corresponds to
\begin{equation}\label{covariance_matrix}
    \Sigma_{\mathrm{opt},\mathrm{out}}(\omega)=
    \begin{pmatrix}
        \langle\left|\delta X_{\mathrm{out}}(\omega)\right|^2\rangle & \mathrm{Cov}_{XY}(\omega)
        \\ 
        \mathrm{Cov}_{XY}^*(\omega) & \langle\left|\delta Y_{\mathrm{out}}(\omega)\right|^2\rangle
    \end{pmatrix},
\end{equation}
where $\mathrm{Cov}_{XY}(\omega)=\frac{1}{2}\langle\{\delta X_{\mathrm{out}}(\omega),\delta Y_{\mathrm{out}}(-\omega)\}\rangle$.
In turn, this matrix has the following expression, detailed in the Supplementary Material
\begin{align}\label{covariance_matrix}
\Sigma_{\mathrm{opt},\mathrm{out}}(\omega)
&=
S_{\mathrm{opt}}(\omega)\sigma_{\mathrm{opt},\mathrm{in}}S_{\mathrm{opt}}^\dag(\omega)
+ \mathcal{N}_{\mathrm{add}}(\omega),
\end{align}
where
\begin{align}
\mathcal{N}_{\mathrm{add}}(\omega)&=
\begin{pmatrix}
    S_P(\omega)
    \\
    S_{\mathrm{car}}(\omega)
\end{pmatrix}^\transp
\begin{pmatrix}
\sigma_P & \Sigma_{P,\mathrm{car}}
\\
\Sigma^\dag_{P,\mathrm{car}} & \sigma_{\mathrm{car}}
\end{pmatrix}
\begin{pmatrix}
S^\dag_P(\omega)
\\
S_{\mathrm{car}}^\dag(\omega)
\end{pmatrix}
\end{align}
and $\Sigma_{P,\mathrm{car}}=\frac{1}{2}\langle (\mathcal{N}_P \mathcal{N}_{\mathrm{car}}+(\mathcal{N}_{P}\mathcal{N}_{\mathrm{car}})^T\rangle$ is the cross-correlation between the polarization noise and carrier noise. In turn, $\sigma_p$ is the covariance of the polarization noise, while $\sigma_{car}$ is associated to the carriers (see Supplementary Material).

\section{Results for squeezing maps and curves}
Once we compute the covariance matrix of the system analytically, we are able to characterise the output quadratures' spectra completely. 

As a matter of fact, the generalized output quadrature variations are $\delta X_\theta^{\mathrm{out}} = \frac{1}{2}\left(\delta a_{\mathrm{out}} e^{-i\theta} + \delta a^\dagger_{\mathrm{out}} e^{i\theta} \right)$ is identified by $Q(\theta)R_{\mathrm{opt},\mathrm{out}}$, with an appropriate column vector of the rotation matrix $Q(\theta)=(\cos(\theta),\sin(\theta))$,
with its associated variance 
\begin{equation}\label{generalized_variance_spectrum}
     V(\theta,\omega)=
     Q(\theta)\,\Sigma_{\mathrm{opt},\mathrm{out}}(\omega)\,Q^\transp(\theta),
\end{equation}
namely
\begin{align}\label{generalized_variance}
     &V(\theta,\omega)=\cos(\theta)^2 |\delta X(\omega)|^2+\sin(\theta)^2 |\delta Y(\omega)|^2\\&+2\sin(\theta)\cos(\theta) \mathrm{Re}(\mathrm{Cov}_{XY}(\omega))\nonumber.
\end{align}

\begin{figure}[t]
    \centering
    \includegraphics[width=0.9\columnwidth]{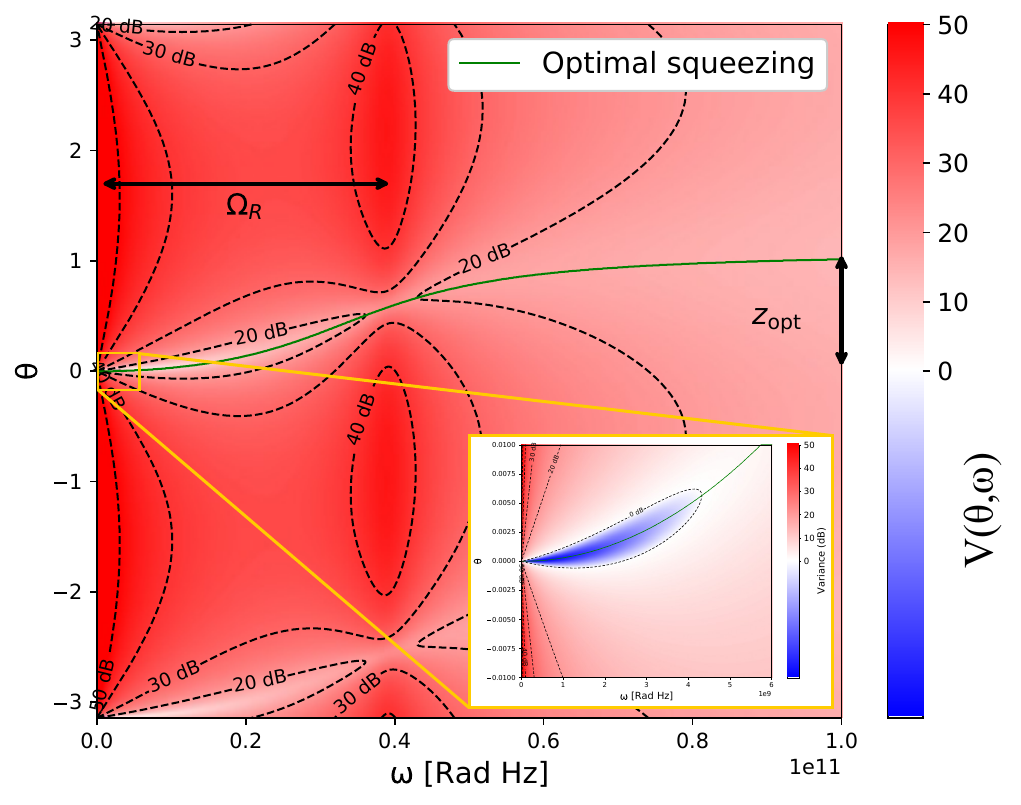}
    \hspace{0.02\columnwidth}
    \includegraphics[width=0.9\columnwidth]{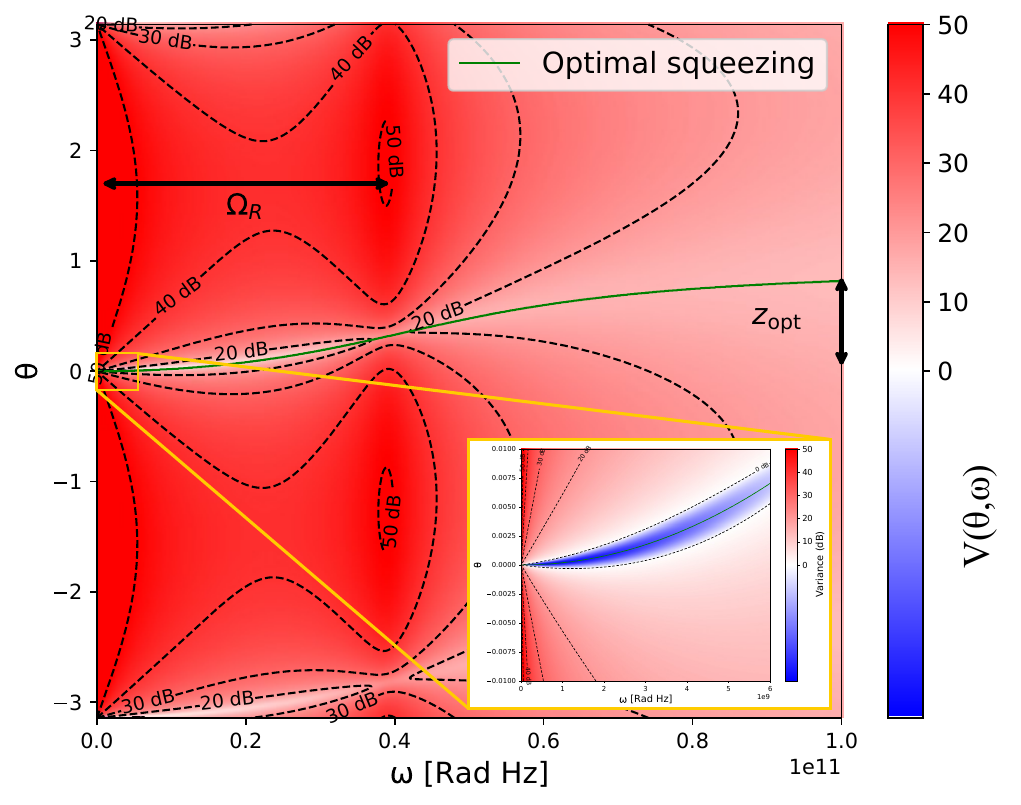}
    \caption{Squeezing maps of $V(\theta,\omega)$ at $a=9$ (a) for $\alpha_H=1,5$ and (b) for $\alpha_H=3$. The role of the $\alpha_H$ factor, while not changing the entity of the amplitude squeezing, is to modify the shape of the optimal squeezing curves (Eq. \ref{optimal_squeezing}) and its asymptotic value $z_{\mathrm{opt}}$, as well as extending the region of squeezing along the optimal curve.}
    \label{squeezing_maps}
\end{figure}

By plotting this quantity (Fig. \ref{squeezing_maps}), we are able to obtain the squeezing maps for the realistic physical parameters of the quantum well from \cite{coldren}, which highlight the quadrature variance spectrum for each value of $\theta$ of the generalised quadrature. This corresponds to the local oscillator phase in the homodyne detection measurement scheme \cite{homodyne}. In the quiet pumping regime, where the noise due to electronic pumping is negligible, and for sufficiently high pumping current, we retrieve a region of squeezing at low frequencies around $\theta=0$. In both cases, the pole at $\omega=0$ of the $Y$ quadrature component, corresponding to $\theta=\pi/2$, determines the divergence of this line \cite{davidovich}. This is a well-known consequence of the phase diffusion process, a characteristic feature of laser dynamics. Apart from that, a second maximum is present around the relaxation oscillations frequency $\Omega_R$. 

It is particularly interesting to investigate the effect of the $\alpha_H$ factor, as it doesn't alter the amplitude squeezing curve, corresponding to $\delta X$, but it determines a shift of the relaxation oscillations maximum on the $\theta$ axis, changing in turn the shape of the optimal squeezing curve. This optimal squeezing curve can be retrieved from the analytical form of the variance of a generalized quadrature in our single-mode scenario from Eq. \ref{generalized_variance}, which we plot in Fig. \ref{squeezing_maps}. It is easy to show that the greatest degree of quantum noise reduction is achieved along the curve parametrized by \cite{davidovich}
\begin{equation}\label{optimal_squeezing}
    \theta(\omega)=\frac{1}{2}\arctan\bigg(\frac{2 \mathrm{Re}(\mathrm{Cov}_{XY}(\omega))}{|\delta X(\omega)|^2-|\delta Y(\omega)|^2}\bigg)
\end{equation}

\begin{figure}[t]
    \centering
    \includegraphics[width=0.90\columnwidth]{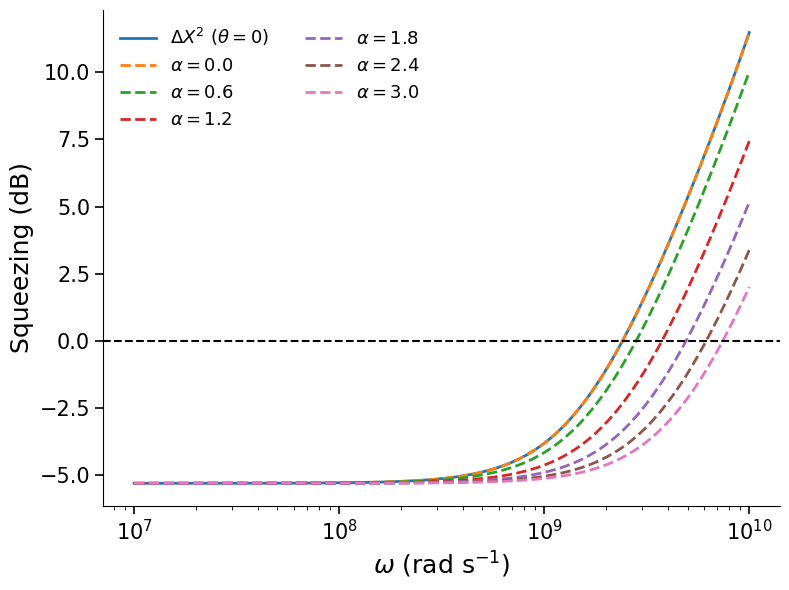}
    \caption{Comparison of the spectrum of amplitude squeezing $\Delta X^2(\omega)$ at low frequencies with the spectrum along the curves of optimal squeezing of Fig. \ref{squeezing_maps}, while varying $\alpha_H$, for $a=9$. In this way, we find the particular role of $\alpha_H$ is to shift the bandwidth over which squeezing is observed.}
    \label{optimalvsdX0}
\end{figure}

If now we plot the amount of squeezing one can obtain when following in line these optimal curves, i.e. choosing the optimal $Q(\theta(\omega))$ for every $\omega$, as in Fig. \ref{optimalvsdX0}, and we compare it with the $X$ quadrature variance, we discover that, by changing the $\alpha_H$ parameter continuously, the region of parameters where squeezing is observed is shifted in frequency, while for zero $\alpha_H$ the amplitude squeezing curve captures all the optimal squeezing. This is a new feature, which does not appear when looking only at the intensity noise spectrum, but is revealed once the complete squeezing map is analysed. 
While the amplitude squeezing at $\omega=0$ corresponds to the global minimum of squeezing, not all the amount of squeezing is accessible when considering the homodyne detection scheme. We find an additional amount of hidden squeezing, only accessible through other measurement schemes. As an example, we propose the synodyne detection scheme \cite{buchmann}, employing a non-monochromatic local oscillator of the form $\alpha(t)=\alpha_- e^{-i\omega_s t}+\alpha_+ e^{i\omega_s t}$. 
In our single mode case, the normalized spinor associated with the measurement scheme can be parametrized as $Q(\theta,\phi)=\begin{pmatrix}\cos(\theta), & e^{i\phi}\sin(\theta)\end{pmatrix}$, as $\alpha_+=\sqrt{I/2}(\cos(\theta)+i\sin(\theta)e^{i\phi})$, $\alpha_-=\sqrt{I/2}(\cos(\theta)+i\sin(\theta)e^{-i\phi})$, where $I=|\alpha_+|^2+|\alpha_-|^2$,
adding a complex phase factor between the two orthogonal quadratures. The resulting variance after the synodyne detection now reads
\begin{equation}
    V(\theta,\phi,\omega)= Q(\theta,\phi)\,\Sigma_{\mathrm{opt},\mathrm{out}}(\omega)\,Q^\transp(\theta,\phi).
\end{equation}

This allows us to have access also to the complex part of the covariance matrix because $V(\theta,\phi,\omega)=\cos(\theta)^2 |\delta X(\omega)|^2+\sin(\theta)^2 |\delta Y(\omega)|^2+2\cos(\phi)\sin(\theta)\cos(\theta) \mathrm{Re}(\mathrm{Cov}_{XY}(\omega))+2\sin(\phi)\sin(\theta)\cos(\theta) \mathrm{Im}(\mathrm{Cov}_{XY}(\omega))$. 

As the synodyne detection scheme can be represented by a projection on a spinor, the result is bounded by the eigenvalues of the complex covariance matrix, the smallest of which is the frequency-dependent minimum of noise, as a consequence of the Rayleigh-Ritz variational principle \cite{Horn2012}, reading
\begin{align}
    &V_{\mathrm{min}}(\omega)=\frac{|\delta X(\omega)|^2+|\delta Y(\omega)|^2}{2}-\\&\sqrt{\frac{(|\delta X(\omega)|^2-|\delta Y(\omega)|^2)^2+|\mathrm{Cov}_{XY}(\omega)|^2}{4}}\nonumber.
\end{align}

\begin{figure}[t]
    \centering
    \includegraphics[width=0.90\columnwidth]{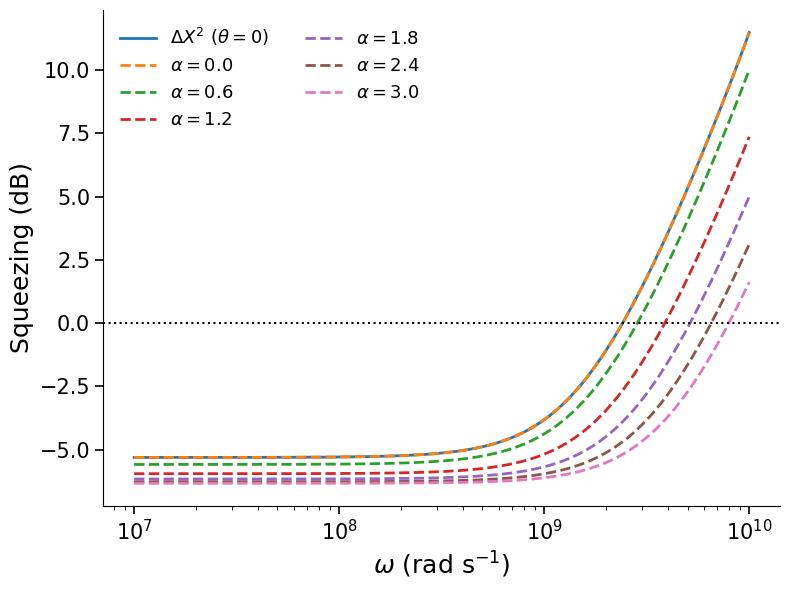}
    \caption{Comparison between the spectrum of amplitude squeezing and the optimal frequency-dependent squeezing with the synodyne detection at $a=9$. In contrast with Fig. \ref{optimalvsdX0}, we observe the global minimum al low frequencies is shifted with $\alpha$, as a further contribution of hidden squeezing is present.}
    \label{Hidden-squeeing}
\end{figure}

\begin{figure}[t]
    \centering
    \includegraphics[width=0.90\columnwidth]{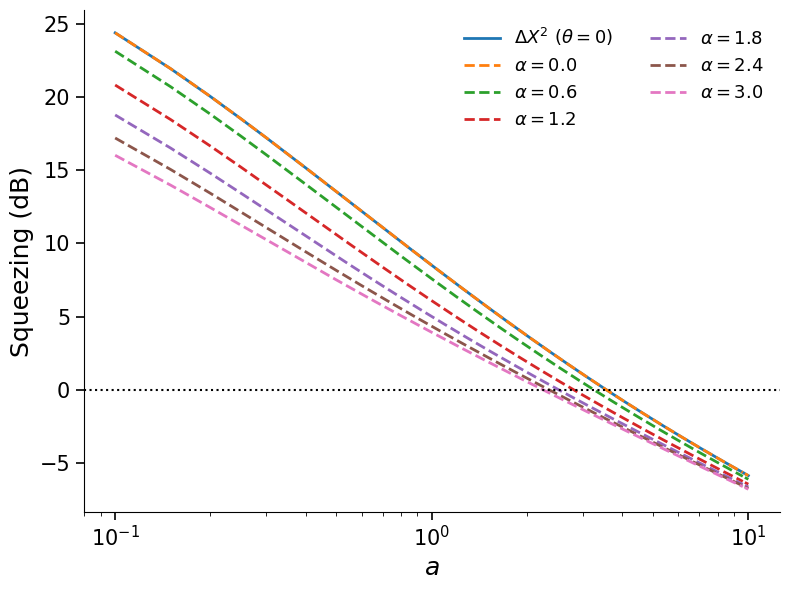}
    \caption{The amplitude squeezing spectrum plotted varying the current a in log scale (here 0 identifies the threshold). As expected, the sub-shot noise region appears sufficiently above threshold after the transition point, where the curve crosses the shot noise (dashed line). An interpolation of the curve reveals a dependence of this quantity in this range as $\approx a^{-3/2}$.
    We compare this quantity with the global minimum of hidden squeezing for different values of $\alpha$. This way, we show the advantage of this global minimum with respect to the minimum of amplitude squeezing in terms of the amount of noise.}
    \label{deltaX20vsa}
\end{figure}

We plot this quantity, highlighting a further contribution of squeezing in the global minimum at $\omega=0$, due to the presence of hidden squeezing (Fig. \ref{Hidden-squeeing}).
Considering only the $\omega \rightarrow 0$ limit, at $\theta=0$ (amplitude squeezing curve, which is also always part of the optimal squeezing curve), one can ask how the amount of squeezing changes with the pumping current and identify precisely what the threshold is for amplitude squeezing to appear (Fig. \ref{deltaX20vsa}). In the range of reasonable values of current that are attainable physically, the amount of squeezing at this point does not surpass 5 dB. Comparing it with the global minimum attainable by leveraging hidden squeezing, we find that the transition point is shifted: squeezing is present at lower current values. Moreover, it's possible to achieve lower noise at the global minimum than in the amplitude noise curve over the considered range of pumping currents.
%

\section{Conclusions and Outlook}
In summary, in this article, we established a fully quantum mechanical framework based on the linearization of Quantum Langevin equations to characterise quadrature squeezing in a semiconductor laser, without resorting to non-rigorous semiclassical methods. Contrary to the case of the atomic laser, treated in \cite{scully} for example, our approach goes beyond the adiabatic elimination of the atomic dynamics, which is not justified for a semiconductor laser, where carriers represent the slowest variables. Furthermore, we took care rigorously of the non-trivial correlations between the optical and the carrier fluctuations.

Rather than considering only the relative intensity noise spectrum, which can also be retrieved in the semiclassical approach \cite{coldren}, the input-output formalism we have employed produces the complete output covariance matrix of the optical output. Thus, we managed to obtain the full squeezing map, which takes into account both the fluctuations in the amplitude and the phase of the laser field. The latter reveals that quiet pumping produces not only amplitude squeezing but also a more complex quadrature squeezing landscape. 

A further relevant aspect emerging from the full quadrature description developed here is the possibility of identifying what can be termed hidden squeezing. In fact, while the covariance matrix fully characterises the Gaussian fluctuations of the output field, its structure in the complex frequency-domain representation reveals that optimal noise reduction may occur along quadratures that are not directly accessible through standard homodyne detection. As a consequence, part of the squeezing resources of the system can remain undetected depending on the measurement scheme, even if they are intrinsically present. We have highlighted that taking into account this hidden squeezing renders possible to lower the amount of noise in the global minimum of squeezing. Beyond its fundamental significance, hidden squeezing represents a potentially exploitable resource for quantum sensing and metrology, as appropriate measurement strategies can be implemented to exploit this improved noise reduction.

This has allowed us to investigate in detail the role of the $\alpha_H$ factor in this physical system, governing the coupling between the carrier dynamics, the intensity and the phase of the field, which is a key feature differentiating the dynamics of semiconductor lasers from atomic/gas lasers. The $\alpha_H$ factor rotates the optimal squeezing curve and modifies the squeezing map, without changing the amplitude squeezing spectrum.

In the present analysis, we focus on the study of a quantum well laser. However, based on this, future work can be undertaken to extend it to other types of semiconductor lasers, such as the quantum dot, which has sparked a renewed interest in applications to quantum technologies for its surprising resilience to the effect of optical feedback compared to the well \cite{Grillot1,Grillot2}. QCLs are of particular interest because their emission range overlaps with the mid-infrared molecular fingerprint region, where the fundamental roto-vibrational transitions of many molecules are located \cite{Faist2013}.

Another important direction for research to follow, drawing on the present article, is to generalise the analysis to a multimodal system. One feature which is of interest in the laser community is the study of quantum correlations between different lines of a spontaneously generated comb, for example, through four-wave mixing in multimode regimes.
\begin{acknowledgments}
The authors would like to acknowledge fruitful discussions with Prof. F. Grillot and Prof. C. Jirauschek. This research was funded, in whole or in part, by Agence Nationale de la Recherche (ANR), project Gandalf ANR-25-CE47-1818.
\end{acknowledgments}
\appendix
\section{Recall of the Hamiltonian description of the quantum well laser as of \cite{ChowKochSargent1994}}
We briefly recall the Hamiltonian description of the quantum well laser, as in the formulation by \cite{ChowKochSargent1994}.

Upon quantization of the electromagnetic medium, the single-mode Hamiltonian reads
\begin{equation}
    H_{ph}=\hbar\Omega\bigg(a^\dag a +\frac{1}{2}\bigg),
\end{equation}
with $\Omega$ being the single-mode frequency of the laser and $a,a^\dag$ the associated creation/annihilation operators.

The carrier part of the Hamiltonian reads
\begin{equation}
    H_{car}=\sum_k H_{car}^k,
\end{equation}
where 
\begin{equation}
    H_{car}^k=\bigg[\epsilon_{g,0}+\frac{\hbar^2 k^2}{2m_e}\bigg]a^\dag_k a_k+\frac{\hbar^2 k^2}{2m_h}b^\dag_{-k} b_{-k},
\end{equation}
where $a_k$ are the operators associated to the electrons in the conduction band and the holes $b_{-k}$ in the valence band with momentum k and $\epsilon_{g,0}$ is the gap energy. The total energy associated to the electron-hole pair is $\omega_k=\epsilon_{g,0}+\frac{\hbar^2 k^2}{2m_e}+\frac{\hbar^2 k^2}{2m_h}$

The interaction term between the carriers and the photons can be written in the rotating-wave approximation as
\begin{equation}
    H_{int}=\hbar\sum_k(g_k^* a^\dag a_k b_{-k}+h.c.).
\end{equation}
This interaction term creates a photons by destroying an electron-hole pair and contains also the inverse process.
\section{Further details about the linearization procedure}
We recall that the system of quantum Langevin equations reported in the Main text for the field creation/annihilation operators and the number of carriers from \cite{ChowKochSargent1994} read
\begin{equation}\label{QLE}
    \begin{cases}        
    \dot{A}=-\frac{1}{2\tau_p}A+\frac{G(N)}{2} A+F_{A} \\
    \dot{N}=\Lambda-R_{sp}-\gamma_{nr}N-g(N)A^\dag A+F_N.
    \end{cases}
\end{equation}
The noise term comprises of two terms
\begin{equation}\label{noise_A}
    F_A(t)=F(t)+F_P(t),
\end{equation}
where the first one is the input noise $F(t)=-\sqrt{\kappa}a_{in}$, while the second one contains the noise associated to the microscopic polarization $F_P$.

We linearize the system around the stationary state
\begin{eqnarray}
    A\simeq (A_0^s+\delta A)e^{i\nu_L t}\\
    N\simeq N_{st}+\delta N.
\end{eqnarray}

First of all, the stationary state, obeying $A_0^s=\sqrt{N_p^{st}}$ reads 
\begin{equation}
    \begin{cases}
     0=-\frac{1}{\tau_p}N_p^{st}+g(N_{st}) N_p^{st}\\
    0=\Lambda-R_{sp}(N_{st})-\gamma_{nr}N_{st}-g(N_{st})N_p^{st}.  
    \end{cases}
\end{equation}
The stationary phase $\nu_L$ obeys
\begin{align}
    &\nu_L=\frac{i}{2}\frac{d}{dt}\bigg[\log\bigg(\frac{\langle A\rangle_s}{\langle A^\dag\rangle_s}\bigg)\bigg]=\frac{i}{2}\bigg\{\frac{\langle \dot{A}\rangle_s}{\langle A\rangle_s}-\frac{\langle \dot{A}^\dag\rangle_s}{\langle A^\dag\rangle_s}\bigg\}\\&=\frac{\alpha_H}{2}g(N_{st})\nonumber.
\end{align}
where $\langle \rangle_s$ represents the average over the noise in the semiclassical equation. This equation can be interpreted as a "renormalization" of the lasing frequency $\Omega$ as a result of the finite alpha factor. The complex linearized correction of the field annihilation operator reads
\begin{align}
    &\frac{d\delta A}{dt}=-\frac{1}{2\tau_p}\delta A+\frac{g(N_{st})}{2}\delta A+\frac{1}{2}(1-i\alpha_H)\frac{\partial g(N)}{\partial N}|_{st}\delta N\\&+\tilde{F}_A \nonumber,
\end{align}
where $\tilde{F}_A=F_A e^{-i\nu_L t}$. In turn the linearized number of carrier $\delta N$
\begin{align}
    &\frac{d\delta N}{dt}=-\gamma_{nr} \delta N -\frac{\partial R_{sp}}{\partial N}\delta N-\frac{\partial g(N_{st})}{\partial N} N_p^{st}\delta N\\&-g(N_{st})A_0^s(\delta A+\delta A^\dag)+F_N\nonumber.
\end{align}
Now, we perform the change of variables to the quadratures with the convention $\delta X=\frac{\delta A+\delta A^\dag}{2}$ and $\delta Y=\frac{\delta A-\delta A^\dag}{2i}$
\begin{align}
    &\frac{d\delta X}{dt}=-\frac{1}{2\tau_p}\delta X+ \frac{g(N_{st})}{2}\delta X+\frac{1}{2}\frac{\partial g(N)}{\partial N}|_{st}\delta N+F_X,
\end{align}
\begin{align}
    &\frac{d\delta Y}{dt}=-\frac{1}{2\tau_p}\delta Y+ \frac{g(N_{st})}{2}\delta Y-\frac{\alpha_H}{2}\frac{\partial g(N)}{\partial N}|_{st}\delta N+F_Y.
\end{align}
Here $F_X$ is defined as $F_X=\frac{\tilde{F}_A+\tilde{F}_A^\dag}{2}$ and $F_Y=\frac{\tilde{F}_A-\tilde{F}_A^\dag}{2i}$. As a consequence, these noise terms can be decomposed as $F_X=-\sqrt{\kappa}X_{in}+F_{X_P}$ and $F_Y=-\sqrt{\kappa}Y_{in}+F_{Y_P}$, where $X_{in}=\frac{a_{in}+ a^\dag_{in}}{2}$ $Y_{in}=\frac{a_{in}- a^\dag_{in}}{2i}$, and $F_{X_P}=\frac{F_P+F_P^\dag}{2}$ $F_{Y_P}=\frac{F_P-F_P^\dag}{2i}$. The equation for $\delta N$ can be cast into the form
\begin{align}
    &\frac{d\delta N}{dt}=-\gamma_{nr} \delta N -\frac{\partial R_{sp}}{\partial N}\delta N-\frac{\partial g(N_{st})}{\partial N} N_p^{st}\delta N\\&-2g(N_{st})A_0^s\delta X+F_N\nonumber.
\end{align}
Together, they constitute the system of Eq. (4), which can then be cast into the form of Eq. (9). 
\section{Detailed description of the noise terms and of their correlations}
As of Eq. (9), the noise part comprises of three terms, namely $\mathcal{N}_{car}$, $\mathcal{N}_P$ and $\mathcal{N}_{opt}$.
Their definition is $\mathcal{N}_{car}=F_N$, $\mathcal{N}_{P}=\begin{pmatrix}
    F_{X_P}\\
    F_{Y_P}
\end{pmatrix},$
and
$\mathcal{N}_{opt}=\begin{pmatrix}
    X_{in}\\
    Y_{in}
\end{pmatrix}$.
The covariance of the optical part, $\sigma_{opt,in}$, appearing in Eq. (18), reads
\begin{align}
    &\sigma_{opt,in}=\frac{1}{2}\langle\begin{pmatrix}
        X_{in}(\omega) \\ Y_{in}(\omega)
    \end{pmatrix}\begin{pmatrix}
        X_{in}(-\omega) & Y_{in}(-\omega)
    \end{pmatrix}\\&+\bigg(\begin{pmatrix}
        X_{in}(-\omega) \\ Y_{in}(-\omega)
    \end{pmatrix}\begin{pmatrix}
        X_{in}(\omega) & Y_{in}(\omega)
    \end{pmatrix}\bigg)^T\rangle
\end{align}
The void input satisfies $\langle a_{in}(\omega)a_{in}^\dag(\omega')\rangle=\frac{\delta(\omega+\omega')}{\sqrt{2\pi}}$ and $\langle a_{in}(\omega) a_{in}(\omega')\rangle=\langle a^\dag_{in}(\omega) a^\dag_{in}(\omega')\rangle=0$. Therefore $\langle X_{in}(\omega) X_{in}(\omega')\rangle=\frac{1}{4\sqrt{2\pi}}\delta(\omega+\omega')$, $\langle Y_{in}(\omega) Y_{in}(\omega')\rangle=\frac{1}{4\sqrt{2\pi}}\delta(\omega+\omega')$, while $\langle X_{in}(\omega) Y_{in}(\omega')+Y_{in}(\omega) X_{in}(\omega')\rangle=0$. In short, the matrix reads
\begin{equation}
    \sigma_{\mathrm{opt,in}}=\frac{1}{4\sqrt{2\pi}}\mathbf{I}_2
\end{equation}
The scalar $\sigma_{car}=\langle F_N(\omega) F_N(\omega') \rangle=2\langle D_{NN}\rangle \frac{\delta(\omega+\omega')}{\sqrt{2\pi}}$ is the noise associated to the carrier's number \cite{ChowKochSargent1994}, with 
\begin{equation}
    2\langle D_{NN} \rangle=\xi\Lambda+\gamma_{nr}N+R_{sp}+(2R_{sp}'-g(N))N_p,
\end{equation}
evaluated at the stationary state. 

The matrix $\sigma_P$ contains the contribution due to the polarization in the noise of the electromagnetic field $F_A$. Its definition is
\begin{align}
    &\sigma_{P}=\frac{1}{2}\langle\begin{pmatrix}
        F_{X_P}(\omega) \\ F_{Y_P}(\omega)
    \end{pmatrix}\begin{pmatrix}
        F_{X_P}(-\omega) & F_{Y_P}(-\omega)
    \end{pmatrix}\\&+\bigg(\begin{pmatrix}
        F_{X_P}(-\omega) \\ F_{Y_P}(-\omega)
    \end{pmatrix}\begin{pmatrix}
        F_{X_P}(\omega) & F_{Y_P}(\omega)
    \end{pmatrix}\bigg)^T\rangle.
\end{align}
The noise $F_P$ satisfies \cite{ChowKochSargent1994} $\langle F_P^\dag(\omega) F_P(\omega')\rangle=R_{sp}\frac{\delta(\omega+\omega')}{\sqrt{2\pi}}$, while $\langle F_P(\omega) F_P^\dag(\omega')\rangle=R_{abs}\frac{\delta(\omega+\omega')}{\sqrt{2\pi}}$, and the other combinations vanish. As a consequence, $\langle F_{X_P} F_{X_P}\rangle=\langle F_{Y_P} F_{Y_P}\rangle=\frac{1}{4}(2 R_{sp}-g(N))\frac{\delta(\omega+\omega')}{\sqrt{2\pi}}$ and $\langle F_{X_P}(\omega)F_{Y_P}(\omega')+F_{Y_P}(\omega)F_{X_P}(\omega') \rangle=0$, then
\begin{equation}
    \sigma_P=\frac{1}{4\sqrt{2\pi}}(2 R_{sp}-g(N))\mathbf{1}_2.
\end{equation}

Finally, we have the cross-correlation term between the carriers and the field noise
\begin{equation}\frac{1}{2}\langle (\mathcal{N}_P \mathcal{N}_{car}+(\mathcal{N}_{P}\mathcal{N}_{car})^T\rangle=\begin{pmatrix}
    -\frac{A_0^s}{2\sqrt{2\pi}}(2 R_{sp}-g(N))\\0
\end{pmatrix},\end{equation}
This relation comes from the cross-correlations $\langle F_N(\omega) F_P(\omega')\rangle=-A_0^s R_{sp}\frac{\delta(\omega+\omega')}{\sqrt{2\pi}}$, $\langle F_N(\omega) F_P^\dag(\omega')\rangle=-A_0^s R_{abs}\frac{\delta(\omega+\omega')}{\sqrt{2\pi}}$, $\langle F_P^\dag(\omega) F_N(\omega')\rangle=-A_0^s R_{sp}\frac{\delta(\omega+\omega')}{\sqrt{2\pi}}$ $\langle F_P(\omega) F_N(\omega')\rangle=-A_0^s R_{abs}\frac{\delta(\omega+\omega')}{\sqrt{2\pi}}$. Consequently, we can write $\frac{1}{2}(\langle F_{X_P}(\omega) F_{N}(\omega')\rangle+\langle F_{N}(\omega) F_{X_P}(\omega')\rangle)=-\frac{A_0^s}{2}(2 R_{sp}-g(N))\frac{\delta(\omega+\omega')}{\sqrt{2\pi}}$ and $\frac{1}{2}(\langle F_{Y_P}(\omega) F_{N}(\omega')\rangle+\langle F_{N}(\omega) F_{Y_P}(\omega')\rangle)=0$.
\section{Solution of Eq. (9) and explicit expression of the scattering spectra}
Starting from the system of Equations (9) in the Main text
\begin{equation}\label{nine}
\frac{d}{dt}\begin{pmatrix}
R_{opt} \\
\hline
R_{car}
\end{pmatrix}=\begin{pmatrix}
        0_{2*2} & \vline B \\ \hline
        C & \vline D
\end{pmatrix}\begin{pmatrix}
R_{opt} \\
\hline
R_{car}
\end{pmatrix}+\begin{pmatrix}
\mathcal{N}_{opt}+\mathcal{N}_{P} \\
\hline
\mathcal{N}_{car}
\end{pmatrix}
\end{equation}
In the Fourier transform
\begin{equation}
\left(
\begin{array}{c|c}
j\omega \, 1_{2\times2} & -B \\ \hline
-C & j\omega - D
\end{array}
\right)\begin{pmatrix}
R_{opt} \\
\hline
R_{car}
\end{pmatrix}=\begin{pmatrix}
\mathcal{N}_{opt}+\mathcal{N}_{P} \\
\hline
\mathcal{N}_{car}
\end{pmatrix}.
\end{equation}
Solving this system of equations and substituting $R_{car}$ into $R_{opt}$
\begin{equation}
\begin{aligned}
R_{\mathrm{opt}}
={}&
\Big[
j\omega \mathbf{1}_2
-
B
\left(
j\omega  - D
\right)^{-1}
C
\Big]^{-1}
\\
&\times
\Big\{
B
\left(
j\omega  - D
\right)^{-1}
\mathcal{N}_{\mathrm{car}}
+
\mathcal{N}_{\mathrm{opt}}
+
\mathcal{N}_{P}
\Big\}.
\end{aligned}
\end{equation}
This expression can be casted in the following expression, which explicits the scattering spectra associated to the different channels.
Now, we use the definition of the input noise
\begin{equation}
    \mathcal{N}_{opt}=-\sqrt{\kappa}R_{opt,in},
\end{equation}
and the input-output relations
\begin{equation}
    R_{opt,out}=R_{opt,in}+\sqrt{\kappa}R_{opt}.
\end{equation}
This leads to the spectral form of the output quadratures
\begin{equation}    R_{opt,out}=S_{car}\mathcal{N}_{car}+S_{P}\mathcal{N}_{P}+S_{opt}\mathcal{R}_{\mathrm{opt,in}},
\end{equation}
\begin{align}  
&S_{opt}=\mathbf{1}_2-\kappa[j\omega\mathbf{1}_2-B(j\omega-D)^{-1}C]^{-1}\\
&S_{car}=\sqrt{\kappa}[j\omega\mathbf{1}_2-B(j\omega-D)^{-1}C]^{-1}B(j\omega-D)^{-1}\\
&S_P=\sqrt{\kappa}[j\omega\mathbf{1}_2-B(j\omega-D)^{-1}C]^{-1}
\end{align}
The explicit expression of the three is
\begin{equation}S_{opt}=\begin{pmatrix}
1-\frac{\kappa}{j\omega+\frac{\gamma_{XN}\gamma_{NX}}{j\omega+\gamma_{NN}}}& 0 \\
-\frac{\kappa\gamma_{YN}\gamma_{NX}}{j\omega[j\omega(j\omega+\gamma_{NN})+\gamma_{NX}\gamma_{XN}]}&1-\frac{\kappa}{j\omega}
\end{pmatrix}    
\end{equation}
\begin{equation}S_{P}=\sqrt{\kappa}\begin{pmatrix}
\frac{1}{j\omega+\frac{\gamma_{XN}\gamma_{NX}}{j\omega+\gamma_{NN}}}& 0 \\
\frac{\gamma_{YN}\gamma_{NX}}{j\omega[j\omega(j\omega+\gamma_{NN})+\gamma_{NX}\gamma_{XN}]}&\frac{1}{j\omega}
\end{pmatrix}    
\end{equation}
\begin{equation}S_{car}=\frac{\sqrt{\kappa}}{j\omega+\gamma_{NN}}\begin{pmatrix}
\frac{\gamma_{XN}}{j\omega+\frac{\gamma_{XN}\gamma_{NX}}{j\omega+\gamma_{NN}}} \\
\frac{\gamma_{YN}\gamma_{NX}\gamma_{XN}}{j\omega[j\omega(j\omega+\gamma_{NN})+\gamma_{NX}\gamma_{XN}]}-\frac{\gamma_{YN}}{j\omega}
\end{pmatrix}    
\end{equation}

\section{ Additional squeezing maps}
We report further squeezing maps of the laser that cover the whole examined range of the current and the alpha factor (Fig. \ref{squeezingmaps}) from the threshold to the edge of the interval of physically reasonable values of the current. This highlights how the region of squeezing appears at $a=4$ and becomes more visible for higher values of the currents. We can clearly see how the peak of relaxation oscillations is shifted in frequency with the current, as it is know in the existing scientific literature \cite{coldren}.

\begin{figure*}[h!]
\centering

\begin{subfigure}{0.32\textwidth}
    \centering
    \includegraphics[width=\textwidth]{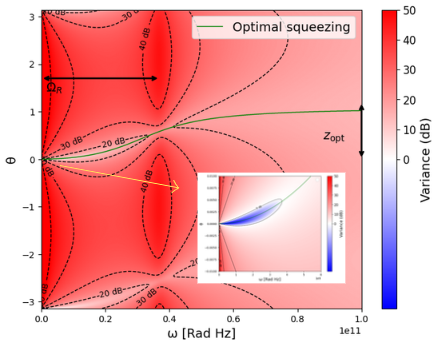}
    \caption{a=8 with $\alpha=1,5$}
\end{subfigure}
\hfill
\begin{subfigure}{0.32\textwidth}
    \centering
    \includegraphics[width=\textwidth]{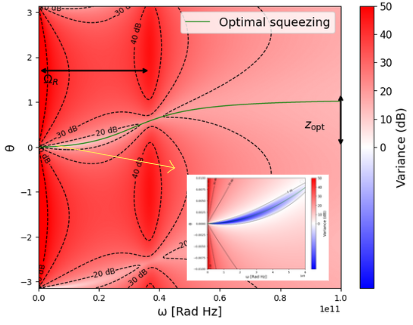}
    \caption{a=8 with $\alpha=3$}
\end{subfigure}
\hfill
\begin{subfigure}{0.32\textwidth}
    \centering
    \includegraphics[width=\textwidth]{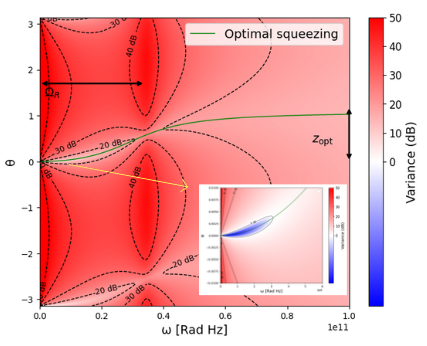}
    \caption{a=7 with $\alpha=1.5$}
\end{subfigure}

\vspace{0.3cm}

\begin{subfigure}{0.32\textwidth}
    \centering
    \includegraphics[width=\textwidth]{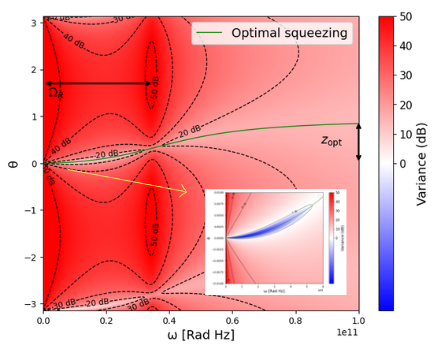}
    \caption{a=7 with $\alpha=3$}
\end{subfigure}
\hfill
\begin{subfigure}{0.32\textwidth}
    \centering
    \includegraphics[width=\textwidth]{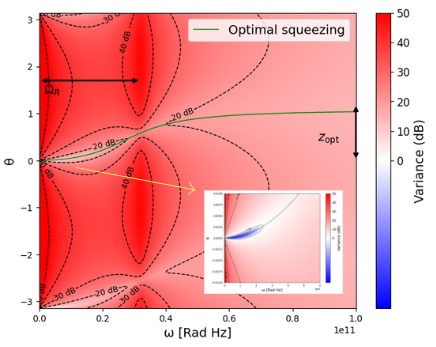}
    \caption{a=6 with $\alpha=1.5$}
\end{subfigure}
\hfill
\begin{subfigure}{0.32\textwidth}
    \centering
    \includegraphics[width=\textwidth]{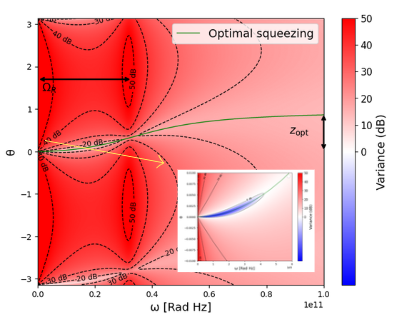}
    \caption{a=6 with $\alpha=3$}
\end{subfigure}
\vspace{0.3cm}

\begin{subfigure}{0.32\textwidth}
    \centering
    \includegraphics[width=\textwidth]{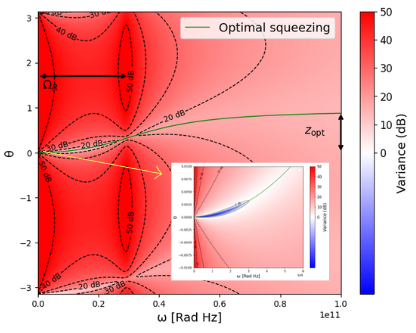}
    \caption{a=5 with $\alpha=3$}
\end{subfigure}
\hfill
\begin{subfigure}{0.32\textwidth}
    \centering
    \includegraphics[width=\textwidth]{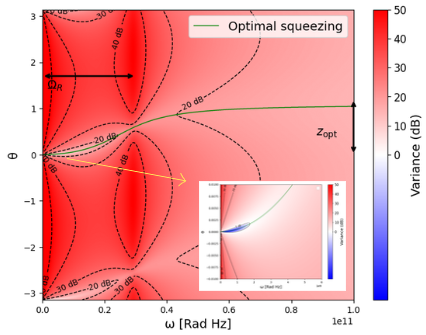}
    \caption{a=5 with $\alpha=1.5$}
\end{subfigure}
\hfill
\begin{subfigure}{0.32\textwidth}
    \centering
    \includegraphics[width=\textwidth]{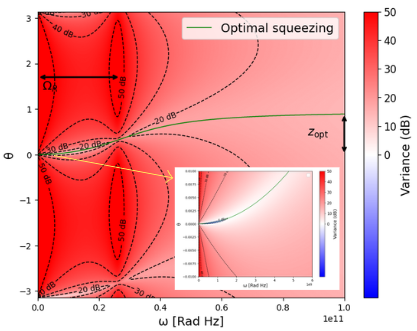}
    \caption{a=4 with $\alpha=3$}
\end{subfigure}
\vspace{0.3cm}

\begin{subfigure}{0.32\textwidth}
    \centering
    \includegraphics[width=\textwidth]{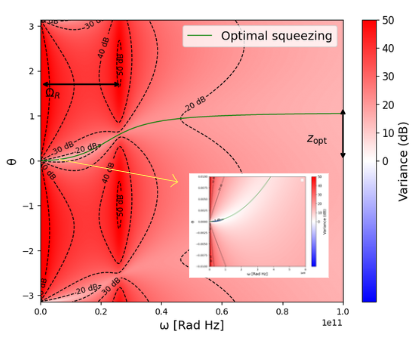}
    \caption{a=4 with $\alpha=1,5$}
\end{subfigure}
\hfill
\begin{subfigure}{0.32\textwidth}
    \centering
    \includegraphics[width=\textwidth]{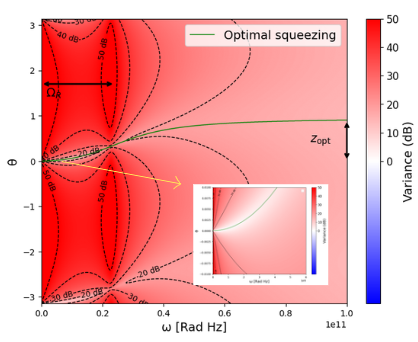}
    \caption{a=3 with $\alpha=3$}
\end{subfigure}
\hfill
\begin{subfigure}{0.32\textwidth}
    \centering
    \includegraphics[width=\textwidth]{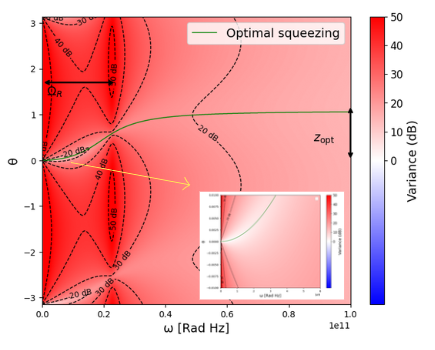}
    \caption{a=3 with $\alpha=1,5$}
\end{subfigure}
\end{figure*}
\begin{figure*}[p]
\ContinuedFloat
\begin{subfigure}{0.32\textwidth}
    \centering
    \includegraphics[width=\textwidth]{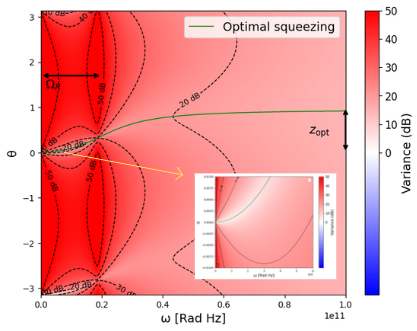}
    \caption{a=2 with $\alpha=3$}
\end{subfigure}
\hfill
\begin{subfigure}{0.32\textwidth}
    \centering
    \includegraphics[width=\textwidth]{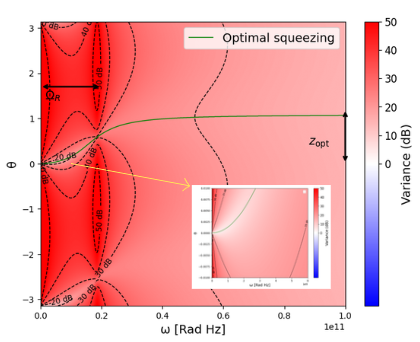}
    \caption{a=2 with $\alpha=1,5$}
\end{subfigure}
\hfill
\begin{subfigure}{0.32\textwidth}
    \centering
    \includegraphics[width=\textwidth]{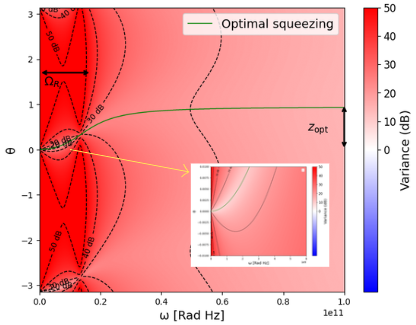}
    \caption{a=1 with $\alpha=3$}
\end{subfigure}
\vspace{0.3cm}
\begin{subfigure}{0.32\textwidth}
    \centering
    \includegraphics[width=\textwidth]{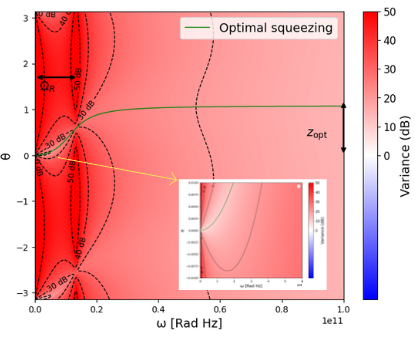}
    \caption{a=1 with $\alpha=1,5$}
\end{subfigure}
\hfill
\begin{subfigure}{0.32\textwidth}
    \centering
    \includegraphics[width=\textwidth]{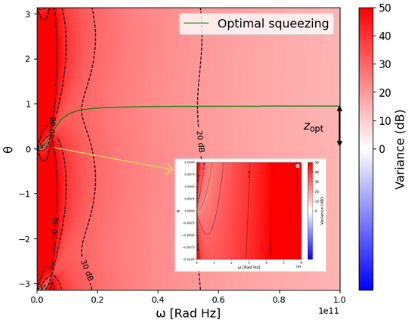}
    \caption{a=0,1 with $\alpha=3$}
\end{subfigure}
\hfill
\begin{subfigure}{0.32\textwidth}
    \centering
    \includegraphics[width=\textwidth]{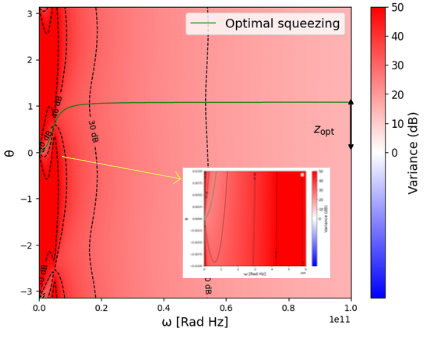}
    \caption{a=0,1 with $\alpha=1,5$}
\end{subfigure}

\caption{Squeezing maps varying the pumping current and the $\alpha_H$ factor. We can observe the shift of the relaxation oscillations with the current, a fact well known in the literature. The region of squeezing appears at $a=4$ and is subsequently enlarged for higher value of the current.}\label{squeezingmaps}
\end{figure*}
\newpage
\bibliography{apssamp}

\end{document}